\documentclass[a4paper,11pt]{article}

\usepackage{amssymb}
\usepackage{amsmath}
\usepackage{tikz}
\usepackage{pgfplots,pgfplotstable}
\usepgfplotslibrary{statistics}
\usepackage{amsthm}
\usepackage{thmtools}
\usepackage{boldline}

\renewcommand{\vec}[1]{\boldsymbol{#1}}
\newcommand{\mat}[1]{\boldsymbol{#1}}

\DeclareSymbolFont{bbold}{U}{bbold}{m}{n}
\DeclareSymbolFontAlphabet{\mathbbold}{bbold}

\newcommand{\R}{\mathbb{R}}
\newcommand{\N}{\mathbb{N}}
\newcommand{\Hm}{\mathcal{H}}

\newcommand{\kernel}{G}

\newcommand{\dotcup}{\ensuremath{\mathaccent\cdot\cup}}
\newcommand{\ps}{X_N}
\newcommand{\A}{\mat{A}}

\makeatletter
\newcommand{\vast}{\bBigg@{3}}
\newcommand{\Vast}{\bBigg@{4}}
\makeatother

\usepackage{algorithm}
\usepackage[noend]{algpseudocode}
\usepackage{float}
\newfloat{algorithm}{t}{lop}

\declaretheorem[style=definition,qed=$\blacksquare$,numberwithin=section]{definition}
\declaretheorem[style=definition,qed=$\blacktriangle$,sibling=definition]{example}

\begin{document}

\author{H.~Harbrecht\thanks{Department for Mathematics and Computer Science, Unversity of Basel, Switzerland}
\and P.~Zaspel\footnotemark[1]}
\title{A scalable $\Hm$-matrix approach for the solution of boundary integral equations on multi-GPU clusters}
\maketitle

\begin{abstract}
In this work, we consider the solution of boundary integral equations
by means of a scalable hierarchical matrix approach on clusters equipped with
graphics hardware, i.e.~graphics processing units (GPUs). 
To this end, we extend our existing single-GPU hierarchical matrix library
\texttt{hmglib} such that it is able to scale on many GPUs and such that it
can be coupled to arbitrary application codes. Using a model GPU implementation
of a boundary element method (BEM) solver,
we are able to achieve more than 67 percent relative parallel
speed-up going from 128 to 1024 GPUs for a model geometry test case with $1.5$
million unknowns and a real-world geometry test case with almost $1.2$ million
unknowns. On 1024 GPUs of the cluster \textit{Titan}, it takes 
less than 6 minutes to solve the $1.5$ million unknowns problem, with 
$5.7$ minutes for the setup phase and $20$ seconds for the iterative 
solver. To the best of the authors' knowledge, we here discuss
the first fully GPU-based distributed-memory parallel hierarchical
matrix Open Source library using the traditional $\Hm$-matrix format
and adaptive cross approximation with an application to BEM problems.
\end{abstract}

\section{Introduction}
The numerical solution of boundary integral equations 
is an important task in applications from science and
engineering such as electric field computations, 
electromagnetism, acoustic scattering, or fluid mechanics 
\cite{Hsiao2008}. Boundary integral equations arise 
typically from the reformulation of 
\textit{boundary value problems} with constant coefficients. 
In many cases, such a reformulation is advantageous, since 
a discretization of a boundary integral equation requires the 
introduction of degrees of freedom just \textit{on the boundary}, 
while the original partial differential equation needs to be discretized 
\textit{in the full domain}, which might be unbounded in case of exterior 
boundary value problems.

Besides the collocation method, the standard approach 
for discretizing and solving boundary integral equations is based 
on a Galerkin discretization which amounts to the \textit{boundary 
element method (BEM)}, see \cite{Gauletal2003,Sauter2011,
Steinbach2008} for example. BEM applies standard techniques 
known from the finite element method (FEM) to the boundary integral 
equation case. Since the kernel of a boundary integral operator is 
usually not compactly supported and singular at the diagonal, the 
stiffness matrix of a BEM discretization is densely populated and 
computationally expensive to compute. To overcome the cubic 
complexity of direct factorization approaches, iterative solvers with 
fast approximate matrix-vector products are used to solve the 
system of linear equations. Candidates for matrix approximations 
are the panel clustering \cite{Hackbusch1989}, the fast multipole 
method \cite{Greengard1997}, hierarchical ($\Hm$) matrices
\cite{Bebendorf2008,Boerm2003,Hackbusch2015,Hackbusch2016} 
or $\Hm^2$ matrices \cite{Boerm2004,Hackbusch2002,Hackbusch2000}. 
We here focus on $\Hm$-matrices, since these are a widely use approach 
and, together with adaptive cross approximation (ACA) \cite{Bebendorf2003}, 
allow for a purely algebraic construction of the matrix approximation, 
facilitating its use in real-world applications. For a given fixed accuracy, 
it is possible to show that the approximate matrix-vector product of 
$\Hm$-matrices can be done in $O(N \log N)$ operations.

The objective of this work is to solve large-scale BEM problems by the 
hierarchical matrix approach. In fact, we aim for solving systems of 
linear equations from BEM discretizations with hundreds of thousands 
or millions of unknowns. Such problem sizes arise if the underlying 
geometry is either very complex or if a solution with a small 
numerical error is required.

We observe two difficulties when it comes to the solution of large-scale 
BEM problems. First, the computational runtime becomes excessively 
large. Second, the required memory is considerable. While the first 
problem can be addressed by a parallelization of a hierarchical 
matrix library on a single compute node with high amounts 
of memory, the second problem can only be fixed by a 
\textit{distributed-memory} parallelization of the hierarchical matrix 
approach. Our conclusion is to work on a distributed-memory parallel 
implementation for $\Hm$-matrices. In particular, we apply and extend 
the many-core parallel Open Source library \texttt{hmglib} \cite{Zaspel2018,hmglib} 
in order to treat BEM-type problems in a distributed-memory parallel way.

\texttt{hmglib} is a many-core parallel library allowing to set up and apply $\Hm$-matrices on a single \textit{graphics processing unit} (GPU). It has been originally developed in the context of the approximation of system matrices from kernel collocation or kernel ridge regression, where it showed decent performance improvements over a parallel $\Hm$-matrix implementation on standard processors (CPUs). As part of the present work, \texttt{hmglib} has been extended such that it can be applied to arbitrary application codes with dense system matrices, as long as the codes provide a means to evaluate matrix entries and the geometric location of the involved degrees of freedom. That is, \texttt{hmglib} now provides a general interface e.g.~for BEM codes. Moreover, and much more important, the library has been extended such that it is now able to run on a distributed-memory parallel cluster of GPUs. This allows to scale the solution of BEM problems on up to millions of unknowns. While the original implementation of \texttt{hmglib} in \cite{Zaspel2018,hmglib} could only pre-compute low-rank blocks, the new implementation is further able to pre-compute and store the (non-admissible) dense matrix blocks in (GPU) memory. This is crucial for BEM applications.

Note that there is a series of related CPU libraries for parallel 
hierarchical matrices. $\Hm$-$\mbox{\texttt{Lib}}^{\mbox{\texttt{pro}}}$ 
\cite{Boerm2003,Grasedyck2008,Kriemann2005,Kriemann2017} is a 
commercial shared-memory parallel library with limited distributed-memory 
support. \texttt{AHMED} (Another software library on hierarchical matrices 
for elliptic differential equations) \cite{Bebendorf} and \texttt{DMHM} 
(Distributed-Memory Hierarchical Matrices) \cite{Poulson} provide 
distributed-memory support on CPUs. \texttt{H2Lib} \cite{Boerm2017} 
is shared-memory parallel. In the related field of Hierarchically 
Semi-Separable (HSS) matrices \cite{Sheng2007}, the software 
\texttt{STRUMPACK} \cite{Ghysels2016,Rouet2016} is shared- and 
distributed-memory parallel. In context of many-core processors, i.e.~GPUs 
and e.g.~Intel Xeon Phi, there is some recent work. An extension to the
\texttt{H2Lib} \cite{Boerm2015} allows to accelerate the quadrature
in a $\Hm^2$ matrix method for BEM by GPUs. A similar approach
is used in the BEM library \texttt{Bempp} \cite{Smigaj2015,Vater2017}.
Furthermore, \cite{Kriemann2013} discusses a many-core parallel
LU-factorization for $\Hm$-matrices on a Xeon Phi device. \texttt{BEM4I} 
\cite{Kravcenko2018,Merta2018} provides a BEM library with ACA 
running on clusters of multi-/many-core hardware by Intel
based on an MPI, OpenMP and vectorization parallelization.
In \cite{Ohshima2018}, the $\Hm$-matrix vector product (without setup)
has been parallelized on a single GPU and on Intel processors.
The new \textit{tile low rank} (TLR) format is used in \texttt{HiCMA} a
library for low-rank Cholesky factorizations on clusters of multi-core
and many-core hardware running on Intel hardware \cite{Akbudak2018}
and with the main application of Mat\'ern-type covarice matrices.
By some of the authors of \cite{Akbudak2018}, further work has been
carried out, which focused on batched dense linear algebra kernels 
\cite{Charara2017} on a single GPU, batched QR and SVD algorithms 
\cite{Boukaram2018} on GPUs and a batched TLR GEMM operation 
on a single GPU \cite{Charara2018}. However, to the best of the authors' 
knowledge, we here discuss the first fully GPU-based distributed-memory 
parallel hierarchical matrix Open Source library using the traditional 
$\Hm$-matrix format and adaptive cross approximation being applied 
to BEM problems.

To be able to apply our parallel library to a (large-scale) BEM model problem, 
we further parallelized an existing sequential CPU code for the solution of elliptic 
problems by the single-layer potential ansatz with piecewise constant basis functions 
on GPU and coupled that code to \texttt{hmglib}. Thereby, we will be able to show that 
we can solve large-scale BEM problems in the range of millions of unknowns  
with a descent strong scaling beyond 68 percent strong 
scaling efficiency on 1024 GPUs of \textit{Titan} at Oak Ridge National Lab 
(starting from 128 GPUs due to memory limitations).

The remainder of this work is structured as follows. In Section~\ref{sec:mathBackground}, 
we introduce the mathematical background of boundary integral equations, the boundary 
element method and the hierarchical matrix approach. Section~\ref{sec:implementation} 
briefly reviews the computational details of \texttt{hmglib} and introduces the new general 
application interface, the dense block storage and multi-GPU parallelization. Numerical 
results and parallel scalability studies are discussed for a model application and a model 
code in Section~\ref{sec:results}. We finish by conclusions in Section~\ref{sec:conclusions}.

\section{Mathematical background}\label{sec:mathBackground}
\subsection{Boundary integral equations}
Shall $\Omega\subset\R^d$ with $d=3$ be a Lipschitz domain and 
$\Gamma:=\partial\Omega$ its surface.
We aim at solving boundary integral equations of type
\begin{equation}\label{eq:boundaryIntegralEquation}
(\mathcal{A} u)(\vec{x}) := \int_\Gamma 
\kernel(\vec{x},\vec{x}^\prime) u(\vec{x}^\prime) \mbox{d}\sigma_{\vec{x}^\prime} 
= f(\vec{x})\,,\quad\vec{x}\in\Gamma\,,
\end{equation}
where $\mathcal{A}$ is supposed to be an invertible boundary integral operator. 
We assume that $\mathcal{A}$ is a continuous and elliptic operator of order
$2q$, which means that it maps from $H^q(\Gamma)$ to $H^{-q}(\Gamma)$. 
We require the integral kernel
$\kernel:\Omega\times\Omega\rightarrow\R$ to be \textit{asymptotically smooth}, 
that is, we require to have constants $C_{as1}, C_{as2}\in\R^{>0}$ such that
$$|\partial_{\vec{x}}^{\vec{\alpha}} \partial_{\vec{x}^\prime}^{\vec{\beta}} 
\kernel(\vec{x},\vec{x}^\prime)| \leq C_{as1}
\frac{(|\vec{\alpha}| + |\vec{\beta}|)!}{(C_{as2}\|\vec{x}-\vec{x}^\prime\|)^{|\vec{\alpha}|+|\vec{\beta}|}} 
 |\kernel(\vec{x},\vec{x}^\prime)|$$
for arbitrary $\vec{x},\vec{x}^\prime\in\Omega$ with $\vec{x}\neq\vec{x}^\prime$
and all multi-indices $\vec{\alpha},\vec{\beta}\in\N_0^d$. This choice allows for kernel functions
with singularities at the \textit{diagonal} $\vec{x}=\vec{x}^\prime$, while being smooth away from the diagonal.

\begin{example}\label{ex:laplace}
Boundary integral equations of the above type arise in context 
of the solution of the Laplace equation
$$
\Delta U = 0\ \mbox{in}\ \Omega, \quad U = f\ \mbox{on}\ \Gamma,
$$ 
where $U\in H^1(\Omega)$ is the solution for given Dirichlet data 
$f\in H^{1/2}(\Gamma)$. Since we know the fundamental solution of the 
Laplace operator, we can make the the single-layer potential ansatz
\begin{equation}\label{eq:singleLayerPotential}U(\vec{x}) = \int_\Gamma \frac{u(\vec{x}^\prime)}{4\pi \|\vec{x}-\vec{x}^\prime\|_2}  
\mbox{d}\sigma_{\vec{x}^\prime}= \tilde{\mathcal{S}}u(\vec{x})\,,\quad\vec{x}\in\Omega\,.\end{equation}
That is, we describe the solution of the Laplace equation by means of 
the unknown density $u\in H^{-1/2}(\Gamma)$. Since the single-layer
potential is continuous in the whole space $\mathbb{R}^d$, the density 
is obtained by solving the boundary integral equation
\begin{equation}\label{eq:exampleBoundaryIntegralEquation}(\mathcal{S}u)(\vec{x}) = \int_\Gamma \frac{u(\vec{x}^\prime)}
{4\pi \|\vec{x}-\vec{x}^\prime\|_2} \mbox{d}\sigma_{\vec{x}^\prime}
= f(\vec{x})\,,\quad\vec{x}\in\Gamma\,,\end{equation}
where $\mathcal{S}:H^{-1/2}(\Gamma)\rightarrow H^{1/2}(\Gamma)$ 
is the \textit{single-layer operator} and $f$ the Dirichlet data of the 
Laplace equation. It can be shown that the kernel function 
$\frac{1}{4\pi\|\vec{x}-\vec{x}^\prime\|_2}$ is asymptotically 
smooth and that $\mathcal{S}$ is continuous and continuously
inver\-tible.
\end{example}

\subsection{Galerkin BEM discretization}
To solve \eqref{eq:boundaryIntegralEquation} by the 
boundary element method (BEM), we first bring the equation 
in its variational form: Find ${u}\in V(=H^q(\Gamma))$, such that 
$$
\int_\Gamma \int_\Gamma \kernel(\vec{x},\vec{x}^\prime) u(\vec{x}^\prime) v(\vec{x}) \mbox{d}\sigma_{\vec{x}^\prime}\mbox{d}\sigma_{\vec{x}} = \int_\Gamma f(\vec{x})v(\vec{x}) \mbox{d}\sigma_{\vec{x}}\ \text{for all}\ v\in V\,.
$$
We discretize it by introducing an approximation 
of the boundary $\Gamma$ by surface elements 
$$\mathcal{T}_h:=\{T_1,\ldots, T_{M} \}$$
of size $O(h)$. The elements $T_i$ are usually chosen as 
planar triangles, cf.~\cite{Gauletal2003,Steinbach2008,
Sauter2011}. Nonetheless, parametric representations
of the surface have recently been become quite popular.
Then, $\mathcal{T}_h$ would be a structured quadrangulation 
and $T_i$ a curved quadrangle, cf.~\cite{HR10,MZB15,Doelzetal2018}.

The elements induce a set of nodes
$$X_h := \{\vec{x}_1,\ldots,\vec{x}_N\}\,.$$
We associate to each node $\vec{x}_i$ a locally 
supported piecewise polynomial $\varphi_i$ of 
order $p$ leading to a finite-dimensional trial space 
$$V_h=\{\varphi_1,\ldots,\varphi_N\} \subset V\,.$$
Then, we look for an approximate solution $u_h\in V_h$, 
such that
$$
\int_\Gamma \int_\Gamma \kernel(\vec{x},\vec{x}^\prime) u_h(\vec{x}^\prime) v_h(\vec{x}) \mbox{d}\sigma_{\vec{x}^\prime}\mbox{d}\sigma_{\vec{x}} = \int_\Gamma f(\vec{x})v_h(\vec{x}) \mbox{d}\sigma_{\vec{x}}\ \text{for all}\ v_h\in V_h\,.
$$
With $u_h(\vec{x}):=\sum_{i=1}^N \alpha_i \varphi_i(\vec{x})$, we finally have to solve the dense linear system
\begin{equation}\label{eq:linearSystem}\mat{A}\vec{\alpha} = \vec{f}\end{equation}
with
$$
\mat{A} = [a_{i,j}]_{i,j=1}^N,\quad 
a_{i,j} = \int_{\Gamma}\int_{\Gamma} \kernel(\vec{x},\vec{x}^\prime) 
\varphi_i(\vec{x}^\prime) \varphi_j(\vec{x}^\prime)\mbox{d}\sigma_{\vec{x}^\prime}\mbox{d}\sigma_{\vec{x}}
$$
and
$$\vec{f} = [f_{i}]_{i=1}^N,\quad f_{i} 
= \int_{\Gamma} f(\vec{x}) \varphi_i(\vec{x})\mbox{d}\sigma_{\vec{x}}\,.$$

\subsection{Hierarchical matrices}\label{sec:hMatrices}
We aim at solving \eqref{eq:linearSystem} by an iterative method. To make this tractable for large $N$,
we use an approximate matrix-vector product for the Galerkin system matrix $\mat{A}$. Our choice is to
use the purely algebraic hierarchical matrices \cite{Boerm2003,Hackbusch2015} with adaptive cross
approximation \cite{Bebendorf2009,Bebendorf2003}, leading to $O(N \log N)$ complexity 
for the matrix-vector product, if we fix the approximation tolerance.

Let us briefly consider the approximation of the Galerkin system matrix $\mat{A}$ by hierarchical matrices.
We first introduce the concept of index sets $I:=\{1,\ldots ,N\}$, representing the nodes
$X_h=\{\vec{x}_1,\ldots,\vec{x}_N\}$ and basis functions $V_N=\{\varphi_1,\ldots,\varphi_N\}$ on $\Gamma$.
Thereby we can associate an index tuple $(i,j)$ to a geometric location and to each entry $a_{i,j}$ of the
system matrix $\A$. We will group these index sets into \textit{clusters} $\tau\subset I$ based on
geometrical arguments. The product of two clusters (i.e.~a \textit{block cluster}),
e.g.~$\tau\times\sigma\subset I\times I$, can then be translated to a sub-matrix
$\left. \A\right|_{\tau\times\sigma}$ of our Galerkin system matrix $\A$. 

The core idea of hierarchical matrices relies on the fact that for an asymptotically smooth kernel, the evaluation of 
$\int_{\Gamma}\int_{\Gamma} \kernel(\vec{x},\vec{x}^\prime) 
\varphi_i(\vec{x}^\prime) \varphi_j(\vec{x}^\prime)\mbox{d}\sigma_{\vec{x}^\prime}\mbox{d}\sigma_{\vec{x}}$
for two geometrically well separated basis functions $\varphi_i, \varphi_j$ can be approximated with a
controlled, small error. This can be expanded to the \textit{admissibility}, i.e.~the approximability,
of a whole block cluster (of nodes). A typical admissibility condition for a block cluster $\tau\times\sigma$
is based on the bounding boxes for clusters $\tau$, $\sigma$ (compare e.g.~\cite{Hackbusch2015} for alternatives).
The bounding box of cluster $\tau\subset I$ is
$Q_\tau:=\prod_{i=1}^3\big[a_\tau^{(i)}, b_\tau^{(i)}\big]$
with $a_\tau^{(i)} := \min_{j\in\tau} x_j^{(i)}$, $b_\tau^{(i)} := \max_{j\in\tau} x_j^{(i)}$
and $\vec{x}_j := \big({x}_j^{(1)},{x}_j^{(2)}, {x}_j^{(3)}\big)^\top$. 
Then, we can introduce the admissibility condition
\begin{equation}\label{eq:admissibilityCondition}
\min\left\{\mbox{diam}(Q_\tau), \mbox{diam}(Q_\sigma)\right\} \leq \eta \mbox{dist}(Q_\tau, Q_\sigma)\,,
\end{equation}
where $\eta\in\R^{\geq 0}$ balances convergence and algorithmic complexity and\linebreak $\mbox{diam}(Q_\tau)$
 and $\mbox{dist}(Q_\tau, Q_\sigma)$ are the diameter and the distance of bounding boxes, 
 respectively, cf.~\cite{Boerm2003}.

Clusters $\tau$ shall always collect geometrically close nodes. They are collected in a \textit{cluster tree}
$\mathcal{T}_I=(\mathcal{V}_I,\gamma)$, which imposes a spatial data structure with a hierarchy on $I$ (or $\ps$).
With $\mathcal{V}_I\subset \mathcal{P}(I)$ being the set of nodes in the tree, i.e.~the clusters, $\gamma$
a mapping $\gamma:\mathcal{V}_I\rightarrow \mathcal{P}(\mathcal{V}_I)$ of each cluster to its hierarchical
sub-clusters, a cluster tree is given such that
\begin{description}
\item[(C1)] $\tau\in \mathcal{P}(I)\setminus\{\emptyset\}$, for all $\tau\in \mathcal{V}_I$,
\item[(C2)] $\mbox{root}(\mathcal{T}) = I$,
\item[(C3)] if $\tau\in \mathcal{V}_I$ is a leaf, i.e.~$\gamma(\tau)=\emptyset$, then $|\tau|\leq C_{leaf}$ and
\item[(C4)] if $\tau\in \mathcal{V}_I$ is no leaf, then it has exactly two children $\gamma(\tau)=\{\tau_1,\tau_2\}$ and $\tau={\tau}_1\,\dotcup\,{\tau}_2$.
\end{description}
In \textit{cardinality-based clustering} (CBC), which will be use in this work, we further
impose $|\tau_1| \approx |\tau_2|$ in \textbf{(C4)}.

\begin{algorithm}[t]
  \caption{Algorithm to build a block cluster tree}\label{alg:buildBlockTree}
  \begin{algorithmic}
    \Procedure{build\_block\_cluster\_tree}{$\tau\times\sigma$, $C_{leaf}$}
	\If{$\tau\times\sigma$ is not admissible and $|\tau|>C_{leaf}$ and $|\sigma|>C_{leaf}$}
	\State $\gamma(\tau\times\sigma)\gets \emptyset$
	\For{$\tau^\prime\in\gamma(\tau)$} \Comment{Loop over children in cluster trees.}
		\For{$\sigma^\prime\in\gamma(\sigma)$}
			\State $\gamma(\tau\times\sigma) \gets \gamma(\tau\times\sigma) \cup \{\tau^\prime\times\sigma^\prime\}$ \Comment{Add new child.}
			\State \Call{build\_block\_cluster\_tree}{$\tau^\prime\times\sigma^\prime$, $C_{leaf}$}
		\EndFor 
	\EndFor
	\Else
		\State $\gamma(\tau\times\sigma) \gets \emptyset$ \Comment{$\tau\times\sigma$ becomes leaf.}
	\EndIf 
    \EndProcedure
  \end{algorithmic}
\end{algorithm}

With a given cluster tree, we can introduce the \textit{block cluster tree}
$\mathcal{T}_{I\times I}=(\mathcal{V}_{I\times I}, \gamma, \mu)$, which builds a hierarchy of block clusters
out of the given cluster hierarchy. Here, $\mathcal{V}_{I \times I}$ is the set of nodes / block clusters in
the tree and $\gamma$ maps a block cluster to its children. Algorithm~\ref{alg:buildBlockTree} recursively
defines the block cluster tree and is launched with $\tau\times\sigma = I\times I$. 

The corresponding sub-matries $\left. \A\right|_{\tau\times\sigma}\in\R^{|\tau| \times |\sigma|}$ of
admissible block clusters in $\mathcal{T}_{I\times I}$ are approximated by an
\textit{$\mathcal{R}(k)$ matrix} $\mat{R}_{\tau\times\sigma}\in\R^{|\tau| \times |\sigma|}$, a matrix of maximum rank $k$,
which is defined as
$$\mat{R}_{\tau\times\sigma}=\mat{U}_{\tau\times\sigma} \mat{V}_{\tau\times\sigma}^\top,
\quad \mat{U}_{\tau\times\sigma}\in\R^{|\tau|\times k},
\quad \mat{V}_{\tau\times\sigma}\in\R^{|\sigma|\times k}\,.$$
Matrix-vector products with $\mathcal{R}(k)$ 
matrices $\mat{R}_{\tau\times\sigma}$ have a computational complexity
of $O\left(r\cdot (|\tau|+|\sigma|)\right)$. We will use the algebraic 
\textit{adaptive cross approximation} (ACA) \cite{Bebendorf2003,Bebendorf2009} to approximate sub-matrices
$\left. \A\right|_{\tau\times\sigma}\in\R^{|\tau| \times |\sigma|}$. ACA can be seen
as a pivoted Gauss elimination and constructs a low-rank approximation
by successive rank-one updates.


Given a rank $k\in\N$ and a block cluster tree $\mathcal{T}_{I\times I}$, we introduce an \textit{$\Hm$-matrix} of
block-wise rank $k$ as matrix $\mat{L}\in\R^{|I|\times |I|}$ such that 
$$\mbox{rank}(\left. \mat{L}\right|_{\tau\times\sigma}) \leq k$$
for all admissible $\tau\times\sigma$. The construction of an $\Hm$-matrix for a dense matrix is known
as \textit{truncation}. Matrix-vector products between $\Hm$-matrices an vectors are realized by a recursive
traversal of the block cluster tree. In each non-admissible leaf, the corresponding (precomputed) full sub-matrix
is applied, while in admissible leafs the (pre-computed) low-rank approximation is applied.
It has been shown, that specific versions of this approach allow to perform an $\Hm$-matrix-vector product
 in complexity $O(k \cdot N \log{N})$ \cite{Hackbusch2015}.


\section{Scalable parallel $\Hm$-matrix approach for BEM}\label{sec:implementation}
Our approximation of the Galerkin matrix by hierarchical matrices is 
based on the library \texttt{hmglib} \cite{Zaspel2018,hmglib}. 
The present work aims at extending \texttt{hmglib} such that it can 
be used for the multi-GPU parallel solution of large-scale boundary 
integral equations discretized by the boundary element method. 
To this end, an abstract code interface, precomputation of dense
matrix blocks and a distributed-memory parallelization had to be 
introduced. In the following, we give a brief overview of the original
implementation \cite{Zaspel2018} and discuss the new techniques 
that have been added to \texttt{hmglib}. 

\paragraph{Remark on technical details.}
Note that an in-depth technical description of state-of-the-art 
GPU-parallel codes requires a lot of technical details, such as 
memory hierarchies, parallelization models, scheduling, caching, 
etc. As in \cite{Zaspel2018}, we here stick to a much less technically 
overwhelming discussion. To this end, we categorize parallel work 
loads either into work loads that can be handled by the use of a 
large amount of parallel threads working on independent tasks or 
into work loads that require the use of more complicated algorithms 
with complex thread interactions, such as reduction operations.
While the first type of work loads can be easily parallelized by 
standard GPU parallelization techniques, i.e.~in CUDA kernels,
we use existing GPU libraries for the second type of work loads, 
whenever this is possible. 

\subsection{\texttt{hmglib} - A many-core parallel $\Hm$-matrix library}
The Open Source GPU library \texttt{hmglib} has originally been 
developed for the approximation of matrices from \textit{kernel interpolation 
/ collocation} or \textit{kernel ridge regression}. It uses a given single 
GPU for all tasks involved in the construction and application of a 
hierarchical matrix, i.e.~it is not an accelerated but a solely GPU-based 
software. To be able to get high performance on GPU, the library
uses a parallel traversal of the block cluster tree, space-filling curves (to build the
clustering) and the concept of {batching for many small similarly-sized tasks}. In 
terms of software and hardware, it requires an \textit{Nvidia} GPU and uses the 
\texttt{CUDA Toolkit}, i.e.~CUDA kernels \cite{Halfhill2008} 
for direct GPU programming, the STL-type algorithm library \texttt{Thrust} 
\cite{Hoberock2010} running in parallel on a GPU and the BLAS/LAPACK-type libraries 
\texttt{CUBLAS} and \texttt{Magma} \cite{dghklty14,hdtld15}.

\begin{figure}
\begin{center}
\hspace*{-2em}\scalebox{0.65}{\includegraphics{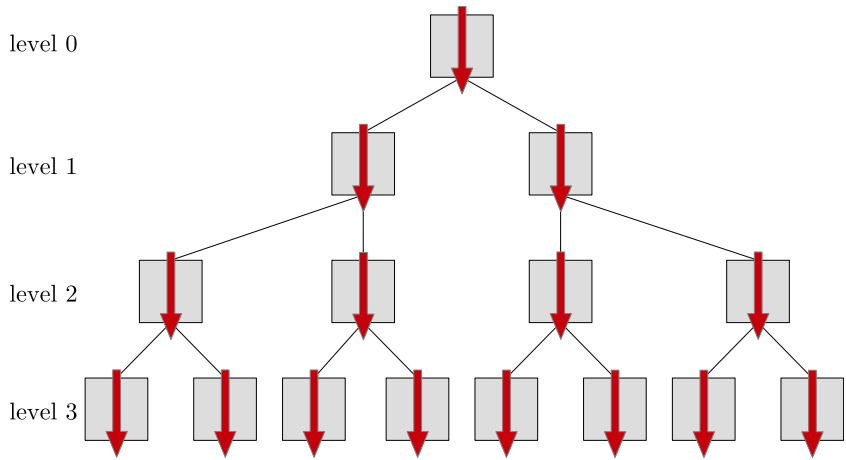}}
\end{center}
\caption{\label{fig:parallelTree}In \texttt{hmglib}, tree traversal is 
parallelized in a level-wise fashion. The red arrows on each level 
correspond to the executed parallel threads.}
\end{figure}

\paragraph{Block cluster tree traversal.}
To get a high parallel performance on many-core hardware, it is necessary 
to express an existing algorithm in a very parallel way. In \cite{Zaspel2018}, 
this has been achieved for the construction and traversal of the block cluster 
by \textit{level-wise} parallelization of the tree traversal. That is, all entries of 
a given level of a tree are computed in a many-core parallel fashion, while 
calculations of offsets in the memory are computed by appropriate parallel 
\textit{scan} operations. Figure~\ref{fig:parallelTree} outlines this methodology. 
The red arrows on a given level correspond to the parallel threads that are 
executed on the GPU. As it becomes obvious, the first few levels of a tree do
not lead to a full parallel utilization. However, on higher levels, this limitation 
is no longer present.

\paragraph{Spatial data structure.}
While classical implementations of $\Hm$-matrices use spatial 
data structures such as \textit{kD-trees} or \textit{quad-/oct-trees}, 
clustering in \texttt{hmglib} is based on \textit{space filling curves} 
\cite{Lauterbach2009,Morton1966,Zaspel2018}. In particular, a 
\textit{Morton code} \cite{Morton1966} is computed for each node 
in $X_h$. This computation is done in a many-core parallel way. 
After sorting the points in $X_h$ following their Morton codes by 
a GPU-parallel sorting method, two consecutive nodes in the 
resulting (sorted) array of nodes are geometrically close. In particular, 
cardinality-based clustering, cf.~Section~\ref{sec:hMatrices},
can be reduced to simple array decompositions.

\begin{figure}
\begin{center}
\scalebox{0.65}{\includegraphics{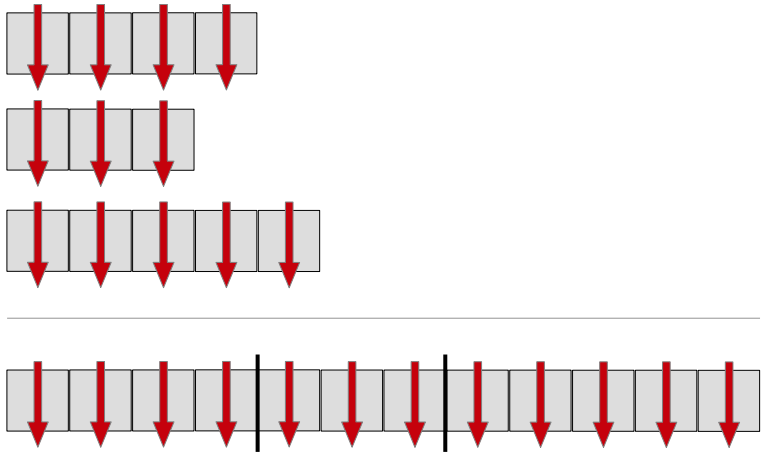}}
\end{center}
\caption{\label{fig:batching}Instead of solving several smaller problems in parallel, \textit{batching},
as used in \texttt{hmglib}, aims at combining all problems into a single much bigger work load, leading to
a higher utilization and higher performance of a GPU.}
\end{figure}

\paragraph{Batching.}
As shown in \cite{Zaspel2018}, the highest impact on the 
GPU-parallel performance of the \texttt{hmglib} code is achieved 
by \textit{batching} of small similarly sized compute tasks into bigger 
batches of compute work. This is specifically used in \texttt{hmglib} 
in the context of the computation of dense matrix blocks 
$\mat{A}_{\tau\times\sigma}$ and low-rank matrix blocks 
$\mat{R}_{\tau\times\sigma}$ and in context of the determination 
of bounding box sizes for the clusters. Figure~\ref{fig:batching} 
outlines the general strategy. Instead of solving (in parallel) 
several small problems (e.g.~the summation of several numbers), 
all these operations are batched together in one bigger work load. 
This allows to achieve a much higher utilization of the GPU and, thus,
leads to higher performance. While this strategy is supported in 
\texttt{Thrust} for, e.g., reduction operations by providing index 
arrays marking the sub-workloads, more complex algorithms 
such as the adaptive cross approximation in \texttt{hmglib} had 
to be adapted to use this new strategy.

\subsection{Abstract program interface.}
In \cite{Zaspel2018}, \texttt{hmglib} was just used for system matrices
 from kernel interpolation / collocation or kernel ridge regression.
Such matrices require the evaluation of a very simple kernel function 
$\kernel$, which was hard-coded. The new developments in the context 
of this work start adding an interface for arbitrary application codes that 
provide custom matrix entries of system matrices that shall be approximated. 
Besides of standard configuration options for $\Hm$-matrices, three general 
\textit{inputs} have to be provided by an application code:
\begin{enumerate}
\item 
The sets of nodes $X_1:=\big\{\vec{x}_1^{(1)},\ldots,\vec{x}_{N_1}^{(1)}\big\}$, 
$X_2:=\big\{\vec{x}_1^{(2)},\ldots,\vec{x}_{N_2}^{(2)}\big\}$.
In our BEM application, we have $X_1=X_2=X_h$.
\item Functions $\mbox{idx}_1:X_1\rightarrow\N$, $\mbox{idx}_2:X_2\rightarrow\N$, 
associating to each node an index. This allows to introduce an \texttt{hmglib}- and 
ordering-independent way to identify the nodes in $X_1$ and $X_2$ by the
application. In our application, this is simply the index of the nodes.
\item 
A callback-type function that can be called by \texttt{hmglib} in order to 
evaluate a single entry $a_{i,j}$ of the system matrix $\mat{A}_{X_1\times X_2}$
for which the $\Hm$-matrix shall be constructed. In our application, this 
is the quadrature routine that computes the matrix entry
$$a_{i,j} = \int_{\Gamma}\int_{\Gamma} \kernel(\vec{x},\vec{x}^\prime) 
\varphi_i(\vec{x}^\prime) \varphi_j(\vec{x}^\prime) 
\mbox{d}\sigma_{\vec{x}^\prime} \mbox{d}\sigma_{\vec{x}}\,.
$$
\end{enumerate}

\paragraph{Technical challenge.}
It turned out that developing a generalized way to provide a callback function for the 
matrix entry evaluation is not easy, at least in connection with the 
CUDA programming language extension provided by the \texttt{CUDA Toolkit}
8.0.
While it is technically possible to use function pointers to \textit{device 
functions}, i.e.~functions that are executed on \textit{and} launched from 
GPU, the practical use of these pointers leads to a strongly reduced 
performance. In order to work around this, we provide a purely virtual 
abstract \textit{device} basis class with a function \texttt{get\_matrix\_entry}. 
This class is overwritten by the application that aims to use \texttt{hmglib}. 
At the same time, \texttt{hmglib} uses \textit{dynamic polymorphism} to 
launch \texttt{get\_matrix\_entry} in an application-independent way.

Nonetheless, this solution introduces two difficulties. The first difficulty 
is in the memory management of the interface \textit{device} class. It 
has to be instantiated and destroyed from \textit{device} code. Therefore, 
its use in a library context, in which the interface code is supposed to run 
on CPU, tends to be rather involved. The second difficulty shows up in 
the compile and link process between the library \texttt{hmglib} and the 
application code. Ideally, the aim would be to provide \texttt{hmglib} as 
a shared library. However, in order to be able to use dynamic polymorphism 
on GPUs in the context of CUDA, it is necessary to put the calling \textit{device} 
code, the abstract \textit{device} basis class and the overwriting \textit{device} 
class into the same \textit{compilation unit}. While it is still possible to individually 
compile the \textit{device} code for the calling code, the basis class and the 
overwriting class into \textit{device}-only object files, these object files have 
to be linked by the device code linker before they can be put together with 
the CPU / \textit{host} code. In practice, this breaks the clear distinction of 
library and application code. To the best of the authors' knowledge, there
is currently no alternative to this approach when using GPUs and \textit{CUDA}.

\subsection{Pre-computation of matrix blocks.}
In \cite{Zaspel2018}, some of us discussed the case of $\Hm$-matrix 
approximation for collocation matrices. The computational effort to compute 
the individual system matrix entries in that case was very low. Therefore, it 
was only considered to pre-compute the low-rank factors for the 
$\mathcal{R}(k)$-matrices $\mat{R}_{\tau\times\sigma}\in\R^{|\tau| \times |\sigma|}$, 
in order to avoid their re-calculation during each $\Hm$-matrix vector product. 

In contrast, computing the entries in the Galerkin matrix for the boundary 
element method is very expensive. In this case, it is extremely important 
to further pre-compute and store the dense blocks $\mat{A}_{\tau\times\sigma}$. 
This has been realized in \texttt{hmglib}. The evaluation of the matrix entries 
is done in a batched way and the matrices are stored in GPU memory. To 
efficiently use the available memory, we store the system matrices continuously 
in GPU memory, without any padding. Pointers to the offsets of each matrix 
are passed to the batched matrix-vector product provided by \texttt{Magma}.

\subsection{Distributed-memory parallelization}\label{sec:parallelization}
We aim at a distributed-memory, i.e.~multi-GPU, parallelization for two 
reasons. First, we want to be able to solve large problems for which
we need a high amount of (GPU) memory. However, GPUs are usually 
rather limited in terms of the available memory. This is why we need 
many GPUs (\textit{scale-up}). Moreover, we want to be able to solve 
BEM problems as fast as possible (\textit{speed-up}).

To fulfill both requirements, we first tried a matrix-based parallelization 
by dividing the large-scale system matrix into blocks of rows, on which 
we independently applied the $\Hm$-matrix approximation. However, 
this lead to a sub-optimal load balancing and sub-optimal speed-up, 
since large admissible matrix blocks were cut into smaller pieces. 
The approach further became prohibitive, as we ran into the situation 
that admissible blocks were divided into skinny (i.e.~wide but thin) 
sub-blocks. Therefore, in the worst case, one of the low-rank factors 
in the ACA could become as large as the number of unknowns in the 
linear system times the required rank, with a strong negative impact on 
scale-up. By introducing a full (row- and column-wise) block-partitioning 
of the system matrix, we could remove this second issue. However, we 
did still cut large admissible blocks into smaller pieces.

\paragraph{Task-based parallelization.}
We overcome most of the mentioned issues by using a \textit{task-based} 
parallelization instead of a matrix-based parallelization. In our task-based
parallelization, we focus specifically on problem size scale-up and 
calculation speed-up in the $\Hm$-matrix construction. This choice is 
valid, since the $\Hm$-matrix construction completely dominates the 
solution process in BEM applications.

\begin{figure}[t]
\begin{center}
\scalebox{0.78}{\includegraphics{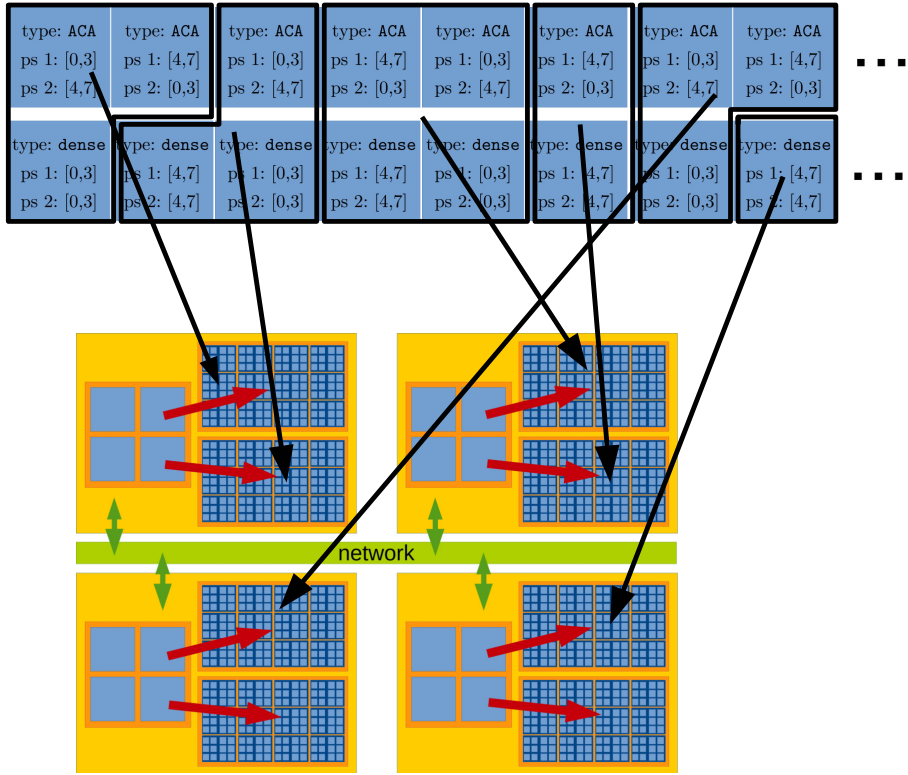}}
\end{center}
\caption{\label{fig:gpuParallelization}The distributed-memory 
parallelization of the $\Hm$-matrix approach distributes similar 
sized subsets of the dense and low-rank matrix blocks to the 
different GPUs. The yellow boxes represent hybrid compute 
nodes in an HPC system that are equipped with a four-core 
CPU and two GPUs.}
\end{figure}

Our task-based parallelization first builds on all GPUs the identical 
global block cluster tree and identifies admissible and non-admissible 
leafs. These leafs are put into two task lists being identical on each 
GPU. Note that no system matrix entry has been evaluated at this stage. 
Then, each of the two task lists is divided into $p$ sub-lists, where $p$ 
is the number of GPU processors. The very computationally expensive 
construction and storage of the low-rank or dense blocks in the sub-lists 
is done in a distributed way on each GPU. 
Figure~\ref{fig:gpuParallelization} illustrates our parallelization concept. 
At the top part of this figure, we show the decomposed task-lists, which 
are then associated to one GPU. Our approach allows to fully decouple 
the construction of the $\Hm$-matrix. In practice, we use the Message 
Passing Interface (MPI) and associate one CPU process / thread to 
one GPU. The only parallelization information, which is required during 
the $\Hm$-matrix construction phase, is the (CPU-)process associated 
to each GPU. It is needed to distribute the sub-lists. The process number 
is provided by MPI.

As part of our current parallelization strategy, we store an identical copy 
of the vector involved in the $\Hm$-matrix-vector product on each GPU.
The $\Hm$-matrix-vector product is applied individually on each GPU. 
That is, it is only performed for those admissible and non-admissible 
blocks that are available on the corresponding processor. To finalize 
the product, we use a global parallel reduction (summation) provided 
by MPI. For improved performance, we rely on a CUDA-aware MPI 
implementation \cite{Kraus2013a}, such that we directly pass pointers 
to GPU memory to the MPI call. Data transfers to and from the network 
adapter are handled by MPI.  

Up to this point, we did not discuss how to partition the task lists into 
sub-lists. This has a strong impact on load balancing. In our current 
implementation, we stick to a rather simplistic scheme. It relies on the 
assumption that the amount of time required for a batched matrix-vector 
product of several dense matrices (or two matrix-vector products of skinny matrices 
in the ACA case) is proportional to the sum over the number of matrix entries 
of all matrices that are involved in the batched product. Concerning the dense 
matrix-vector product task list, we thus balance the storage size of the batched 
matrices on each GPU. Similarly, we balance the storage size of the batched 
low-rank factors for the ACA task list, compare Figure~\ref{fig:gpuParallelization}. 

\paragraph{Scalability and load balancing discussion.}
Independently computing the block cluster tree on each GPU and further 
storing identical copies of the vector involved in the matrix-vector product 
requires, with growing problem size, a growing fixed amount of memory. 
This has a potential impact on the problem size scale-up on GPUs with a 
small amount of GPU memory. We could partially fix this obstruction by 
moving the block cluster tree construction to CPU. Nevertheless, our 
intention is to provide a solely GPU-based implementation. Therefore, 
we do not use the CPU. 

Note, also that 
the currently required global communication in the $\Hm$-matrix-vector 
product might limit the speed-up in the matrix-vector product. We accept 
this, since we right-now focus on calculation speed-up in the $\Hm$-matrix 
construction, as discussed before. Finally, the underlying assumption for 
our task list partitioning strategy strongly depends on the applied batched 
matrix-vector product implementation. Here, we see some room for 
improvements by a more elaborated cost model.

\section{Numerical results}\label{sec:results}
In the following, we will first briefly discuss the model problem and the
applied GPU-based model BEM solver. This is followed by an overview 
of the used hard- and software and the definition of two test cases. The 
first major study of this section is concerned with numerical results that
indicate convergence of the implemented method. The remaining part 
of this section discusses the performance and scalability of the multi-GPU approach.

\subsection{Model BEM solver}
To test our extended version of the \texttt{hmglib} library in the context
of boundary element methods, we have implemented a GPU-based model 
BEM solver. It solves the model problem discussed in Example~\ref{ex:laplace}, 
i.e.~the Laplace problem reformulated by the single-layer potential ansatz and 
resulting in the boundary integral equation~\eqref{eq:exampleBoundaryIntegralEquation}.
We stress here that this BEM solver is solely built with the intention to have a test 
case for the \texttt{hmglib} library in the context of BEM. That is, it is not supposed 
to compete with other BEM libraries.

Our model GPU BEM solver is based on a sequential 
in-house code. This in-house code gets the boundary $\Gamma$ by a 
parametric representation similar to iso-geometric analysis. It discretizes 
$\Gamma$ by a quadrangular mesh and introduces a finite-dimensional 
trial space with piecewise constant ansatz functions. We 
identify basis functions with element centers. In the
Galerkin matrix assembly, higher-order quadrature and the
\textit{Duffy trick} \cite{Duffy1982,Sauter1997} are applied to get
an accurate approximation of the integrals. The resulting system of
linear equations is solved by a conjugate gradient (CG) solver.

In our GPU version of the CPU code, we re-use the existing sequential
CPU code to build the required data structures (mesh, element
lists, \ldots). This data is copied to GPU. Then, the actual matrix
assembly is done on GPU. To this end, we parallelize the fully decoupled 
node-wise assembly operation 
by appropriate CUDA device functions that overwrite
the abstract matrix assembly class of \texttt{hmglib}. As e.g.~discussed
in \cite{Boerm2015}, the complicated, memory-intensive and node-wise sequential
quadrature routines easily lead to a limited GPU \textit{utilization}.
In fact, we had to limit the size of the so-called \textit{thread blocks},
i.e.~the number of threads executed on a symmetric multiprocessor of a GPU,
to $128$ in order to be able to run the quadrature routines. Typical choices 
for most other applications are $512$ or $1024$. The hand-written GPU-based
CG solver uses the multi-GPU parallel $\Hm$-matrix-vector 
product. The solution of the iterative solver is copied back to CPU, where its 
error is evaluated by the standard error evaluation routines of the sequential 
CPU code.

All results of this work are calculated with the $\Hm$-matrix
parameters $\eta = 1.0$ and $C_{leaf} = 32$. The stopping criterion
of the CG solver is a relative residual of $10^{-8}$.

\subsection{Hardware and software setup}\label{sec:hardwareSoftwareSetup}
To run and benchmark \texttt{hmglib} together with the model GPU BEM
solver, we use the former Top 1 HPC system \textit{Titan} (27 Peta-FLOPS),
located at the Oak Ridge National Lab, US. This is a Cray XK7 cluster equipped
with 18688 compute nodes, which are connected by a Gemini interconnect.
Each compute node contains a 16-core AMD Opteron processor, 32 GB of (CPU)
memory and a Nvidia Tesla K20X GPU (Kepler architecture) with 6 GB of
GPU memory.

We use Titan's default \texttt{gcc} compiler, Cray's \texttt{MPICH}
implementation and Titan's (at time of benchmarking) latest
\texttt{CUDA Toolkit} 7.0 (including the corresponding \texttt{Thrust}
and \texttt{CUBLAS} libraries). In addition, we compile and use
\texttt{Magma} 2.3.0 and \texttt{OpenBLAS} 0.2.20 (as dependency of
\texttt{Magma}). In all compilations, we use the standard optimization
flag \texttt{-O3}. All GPU codes are compiled for the best possible
\textit{Compute Architecture 3.5}. The version of \texttt{hmglib} that
is used to create the results in this work is identical to the commit
\texttt{8b4a4ff} of \texttt{hmglib} on Github \cite{hmglib}.

\subsection{Test cases}
\begin{figure}[t]
\begin{center}\vspace{-2em}
\scalebox{0.16}{\includegraphics{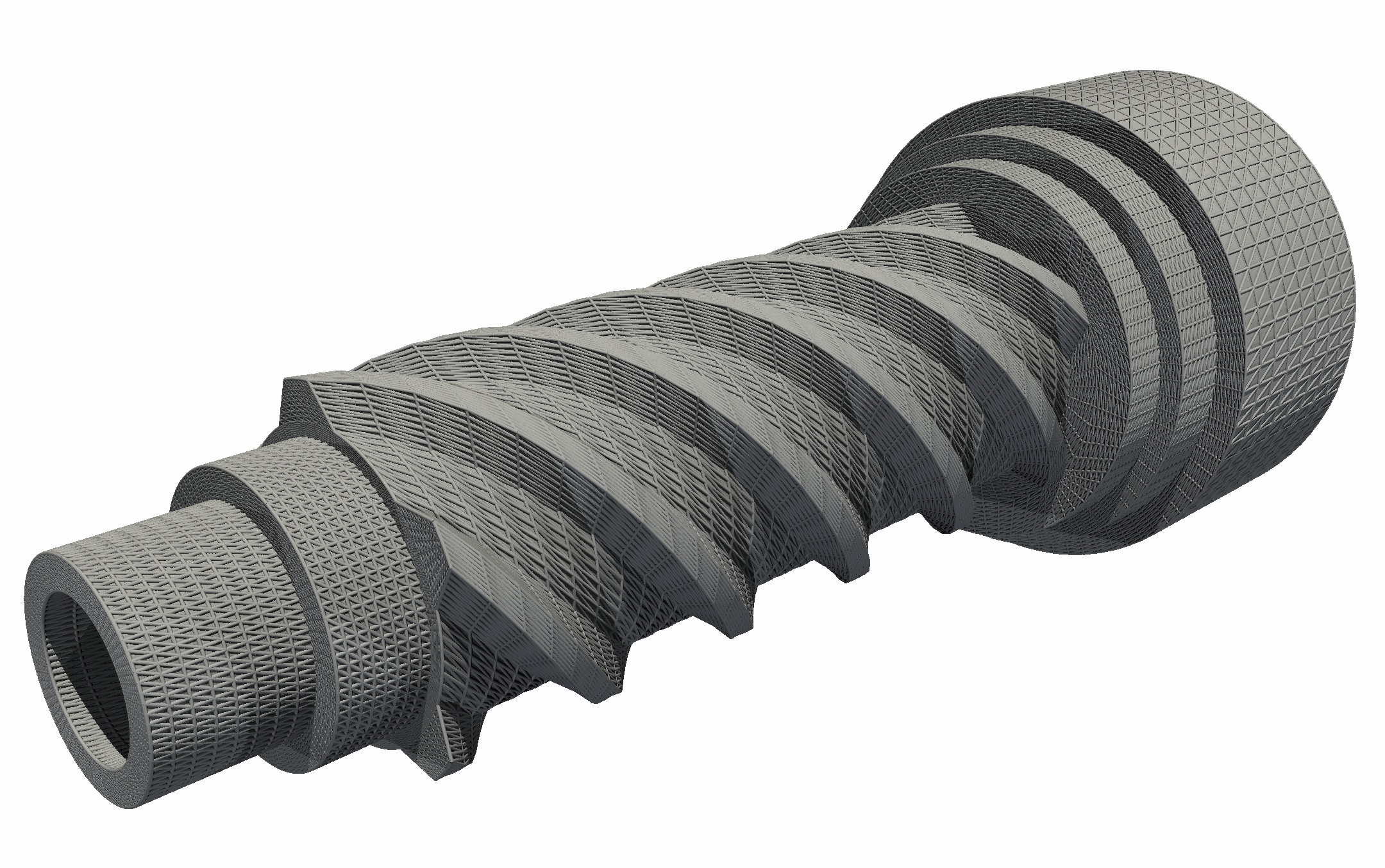}}
\end{center}\vspace*{-2em}
\caption{\label{fig:gearwheel3}As real-world test case, we solve a boundary
integral equation on a complex gearwheel geometry. The overlying mesh
corresponds to a discretization with $N=296960$ boundary elements.}
\end{figure}

To test our implementation, we first use the model geometry $\Omega:=[0,1]^3$,
i.e.~$\Gamma$ is the surface of the unit cube. As real-world test case,
we further consider the geometry of a gearwheel with the bounding box
$[-4,4] \times [-4,4] \times [-11.5,9.1]$, as shown in
Figure~\ref{fig:gearwheel3}. In both cases, the right-hand side $f$
in equation~\eqref{eq:exampleBoundaryIntegralEquation} is 
$$f(\vec{x}):=4{x_1}^2 - 3 {x_2}^2 - {x_3}^2\,.$$
Since the right-hand side is the trace of a harmonic function,
this choice allows us to compare the numerical solution against 
the exact solution $U_{exact}$ of the underlying Laplace equation. 
In particular, we compute the worst-case error
$$
\epsilon(h) := \max_{\vec{x}\in X_{eval}} \left| U_{exact}(\vec{x}) 
- \tilde{\mathcal{S}} u_h(\vec{x})\right|
$$
by evaluating the single-layer potential ansatz from
equation~\eqref{eq:singleLayerPotential} for the approximated
solution $u_h$. Here, $X_{eval}$ is a large set of fixed points in the 
interior of the domain $\Omega$. 

\subsection{Convergence}\label{sec:convergence}
In order to verify the correctness of our GPU BEM model implementation
and of the distributed-memory parallel $\Hm$-matrix implementation, we
first do a classical convergence study, both with respect to the discretization, 
i.e.~the number of boundary elements, and with respect to the
approximation by the $\Hm$-matrix approach.

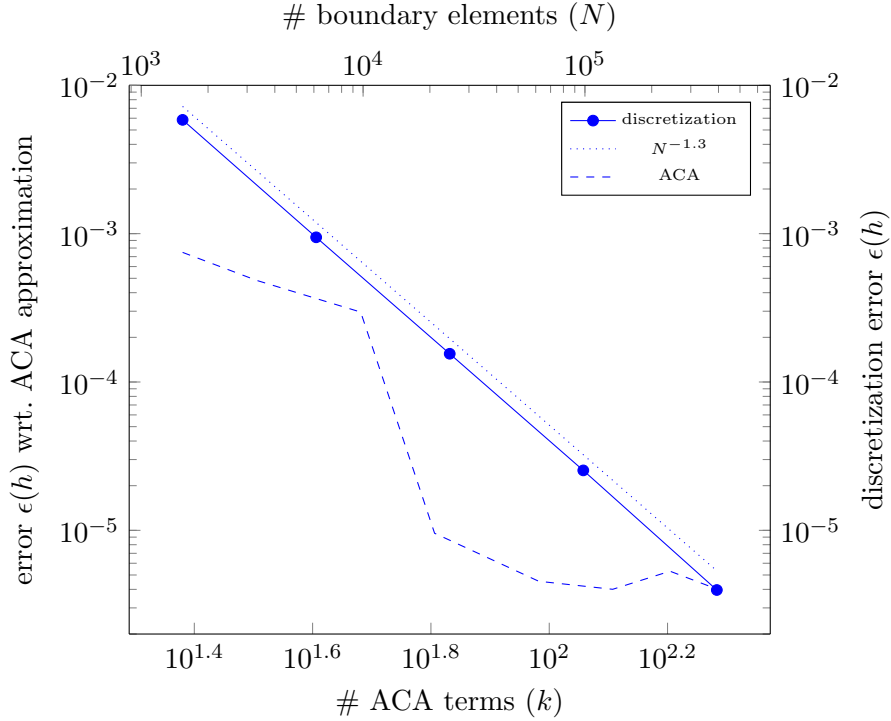
\begin{figure}[t]
\begin{center}\vspace*{-1em}
\begin{tikzpicture}
\begin{loglogaxis}
[xlabel=\# boundary elements ($N$),
ylabel=discretization error $\epsilon(h)$,
xlabel near ticks,
ylabel near ticks,
scale only axis,
ymin=2.0e-6,
ymax=1.0e-2,
axis y line*=right,
axis x line*=top,
skip coords between index={5}{6},
legend style={legend pos = north east, font=\tiny}]
\addplot[blue,mark=*] table{discretization_error_cube.csv};\label{plt:discretizationCube1}
\addlegendentry{discretization};
\addplot[blue,dotted] table[y expr=100/(x^(1.3))] {discretization_error_cube.csv};\label{plt:discretizationCube2}
\addlegendentry{$N^{-1.3}$};
\end{loglogaxis}
\begin{loglogaxis}
[xlabel=\# ACA terms ($k$),
ylabel=error $\epsilon(h)$ wrt.~ACA approximation,
xlabel near ticks,
ylabel near ticks,
scale only axis,
ymin=2.0e-6,
ymax=1.0e-2,
axis y line*=left,
axis x line*=bottom,
legend style={legend pos = north east, font=\tiny}]
\addlegendimage{/pgfplots/refstyle=plt:discretizationCube1}\addlegendentry{discretization}
\addlegendimage{/pgfplots/refstyle=plt:discretizationCube2}\addlegendentry{$N^{-1.3}$}
\addplot[blue,dashed] table{aca_error_cube.csv};
\addlegendentry{ACA};
\end{loglogaxis}
\end{tikzpicture}
\end{center}\vspace*{-1em}
\caption{\label{fig:cubeConvergence}Convergence study for the multi-GPU
implementation with model problem on a cube geometry. The discretization
error \textit{(solid line)} decays with an appropriate algebraic rate
while the error in the ACA \textit{(dashed line)} converges up to exponentially
until it hits the discretization error.}
\end{figure}

\paragraph{Cube geometry.}
We solve the above discussed model problem on the surface of the cube geometry
for a growing number $N$ of boundary elements. In this first experiment, we fix the 
approximation by the adaptive cross approximation to $k=128$. The convergence study 
is done for $N=1536, 6144, 24576, 98304, 393216$. Note that we compute the first two 
results on 4 GPUs and the remaining results on 128 GPUs. Although some of the
problems would already fit in less GPUs (or even one GPU), we keep a higher
number of GPUs to actually check the convergence of the \textit{multi}-GPU code.

Figure~\ref{fig:cubeConvergence} shows the convergence results for this first
test by the solid blue line. The corresponding axes for this test are on the
top and on the right-hand side. We observe an algebraic error decay with a (measured)
rate of $1.3$. If $\Gamma$ would be smooth, we could get a rate of $1.5$.
However, since this is not the case, the observed rate perfectly fits our expectations.
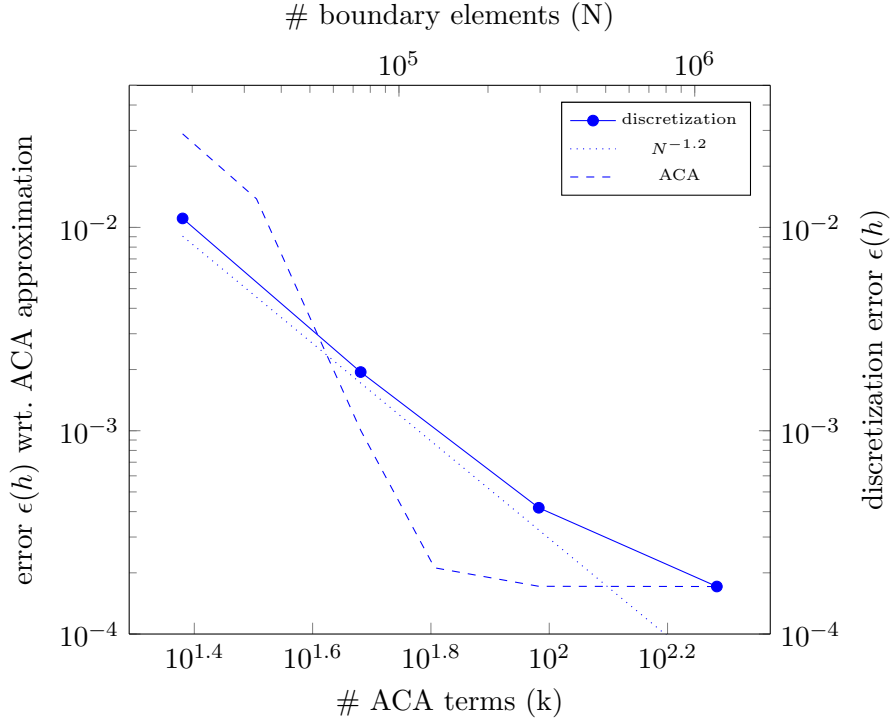
\begin{figure}[t]
\begin{center}\vspace*{-1em}
\begin{tikzpicture}
\begin{loglogaxis}
[xlabel=\# boundary elements (N),
ylabel=discretization error $\epsilon(h)$,
xlabel near ticks,
ylabel near ticks,
scale only axis,
ymin=1.0e-4,
ymax=0.5e-1,
axis y line*=right,
axis x line*=top,
legend style={legend pos = north east, font=\tiny}]
\addplot[blue,mark=*] table{discretization_error_gearwheel3.csv};\label{plt:discretizationGearwheel31}
\addlegendentry{discretization};
\addplot[blue,dotted] table[y expr=1200/(x^(1.2))] {discretization_error_gearwheel3.csv};\label{plt:discretizationGearwheel32}
\addlegendentry{$N^{-1.2}$};
\end{loglogaxis}
\begin{loglogaxis}
[xlabel=\# ACA terms (k),
ylabel=error $\epsilon(h)$ wrt.~ACA approximation,
xlabel near ticks,
ylabel near ticks,
scale only axis,
ymin=1.0e-4,
ymax=0.5e-1,
axis y line*=left,
axis x line*=bottom,
legend style={legend pos = north east, font=\tiny}]
\addlegendimage{/pgfplots/refstyle=plt:discretizationGearwheel31}\addlegendentry{discretization}
\addlegendimage{/pgfplots/refstyle=plt:discretizationGearwheel32}\addlegendentry{$N^{-1.2}$}
\addplot[blue,dashed] table{aca_error_gearwheel3.csv};
\addlegendentry{ACA};
\end{loglogaxis}

\end{tikzpicture}
\end{center}\vspace*{-1em}
\caption{\label{fig:convergenceGearwheel3}Our multi-GPU parallel BEM solver together with the
\texttt{hmglib} library is also applied to the very complex gearwheel geometry. Here it shows
a similar discretization error \textit{(solid line)} and ACA approximation error 
\textit{(dashed line)} as for the unit cube geometry.}
\end{figure}

We further check the convergence of the adaptive cross approximation. To this end, we fix 
the number of boundary elements to $N=393216$ and gradually increase the number of terms
used in the ACA as $k=24, 32, 48, 64$, $96, 128, 160, 192$. In this second test, we always use
128 GPUs. The results are depicted in Figure~\ref{fig:cubeConvergence} as the dashed
line. The error in the ACA decays up to exponentially until it hits the discretization
error for roughly $k=128$. Beyond that, it stagnates with small fluctuations.

\paragraph{Gearwheel geometry.}

Next, we repeat our previous studies with the complex real-world geometry of the gearwheel
seen in Figure~\ref{fig:gearwheel3}. We again fix the low-rank approximation to $k=128$ and
increase the number of boundary elements in accordance with $N=18560, 74240, 296960, 
1187840$. The first two problem sizes are computed on 256 GPUs and the second two 
problem sizes are computed on 1024 GPUs.

The results are given in Figure~\ref{fig:convergenceGearwheel3} by the solid line. It
shows a measured algebraic rate of roughly $1.2$, which fits again our expectations
since the gearwheel geometry is not smooth. For the largest problem size, we get a 
slight degradation in this rate. This might be due to our fixed stopping criterion of a 
relative residual of $10^{-8}$ in the CG solver and a potentially high condition number 
of the corresponding Galerkin system matrix. 

We also repeat the convergence study for ACA for fixed $N=1187840$ 
and growing
$k=24, 32, 48, 64, 96, 128, 160, 192$. As before, we observe an up to exponential
convergence until the discretization error is hit. Beyond that, there are again only
small variations on the same error level.

To summarize, our distributed-memory multi-GPU parallel model BEM code applied
together with \texttt{hmglib} perfectly matches our convergence expectations.
This holds for the model geometry of a unit cube and for the very complex
gearwheel geometry.

\subsection{Performance and scalability}\label{sec:performanceAndScalability}
\begin{table}
\begin{center}
\scalebox{1}{\begin{tabular}{crcc|rr}\hlineB{3}
 & \multicolumn{3}{c|}{} & \multicolumn{2}{c}{runtime}\\\hline
 geometry& $N$ & $k$ & $p$ & $\Hm$-setup & CG solver\\
& & & & \multicolumn{1}{c}{[s]} & \multicolumn{1}{c}{[s/iter.]}\\\hlineB{2}
cube & 1536 & 24 & 1 & 0.86 & 0.0080\\ 
     & 6144 & 24 & 1 & 5.44 & 0.0147\\
     & 24576 & 24 & 1 & 39.91 & 0.0350\\
     & 98304 & 48 & 8 & 163.15 & 0.1504\\
     & 393216 & 48 & 32 & 698.49 & 0.0914 \\
     & 1572864 & 48 & 128 & 1880.26 & 0.1918\\\hline
     & 393216 & 24 & 128 & 46.66 &0.0335 \\
     &  & 32 & 128 &101.01 &0.0284  \\
     &  & 48 & 128 &245.35 &0.0288 \\
     &  & 64 & 128 &342.47 &0.0333 \\
     &  & 128 & 128 &518.79 & 0.0342\\
     &  & 160 & 128 &555.77 & 0.0393\\\hline
     & 1572864 & 48 & 128 & 1832.61 &0.1124 \\
     & & 48 & 256 & 955.92 &0.0858\\
     & & 48 & 512& 543.12 & 0.0767 \\
     & & 48 & 1024 & 338.80 & 0.0867\\\hlineB{2}
gearwheel &  1187840 & 24 & 1024 & 497.41 & 0.0642\\
          &        & 32 & 1024 & 509.11 & 0.0585 \\
          &        & 48 & 1024 & 534.10 & 0.0583 \\
          &        & 64 & 1024 & 684.75 & 0.0600 \\
          &        &128 & 1024 & 864.29 & 0.0643 \\
          &        &160 & 1024 & 877.60 & 0.0783 \\\hline
 & 1187840 & 24 & 128 & 2726.23 & 0.0969\\
 &         & 24 & 256 & 1486.17 & 0.0827\\
 &         & 24 & 512 &  937.07 & 0.0633\\
 &         & 24 &1024 &  497.90 & 0.0616\\\hlineB{3}
\end{tabular}}
\end{center}
\caption{\label{tab:runtimes}The above table collects runtime benchmarks
done with \texttt{hmglib} and our model GPU BEM solver on $p$ GPUs.
We are able to e.g.~solve a BEM problem on a cube geometry with about $1.5$
million boundary elements on GPUs in less than 6 minutes (5.7 minutes for the
$\Hm$-matrix setup, $20$ seconds for the CG solver).}
\end{table}

To assess the properties of our implementation, we performed a series
of benchmarking and scalability studies. All studies were carried
out on \textit{Titan}, cf.~Section~\ref{sec:hardwareSoftwareSetup}.
Time measurements in our distributed-memory multi-GPU parallelization
are wall clock times for the slowest parallel process, i.e.~the slowest
GPU. If not otherwise stated, these worst-case times were averaged over
five runs to reduce the impact of changing loads on the utilized HPC system.
Timings for all measurements are collected in Table~\ref{tab:runtimes}.

\paragraph{Scale-up in problem size.}
We first discuss the scale-up in terms of problem size, exemplified
for the cube geometry. The corresponding timings are given in the first
block of Table~\ref{tab:runtimes}. We are able to solve a problem with
up to 24576 boundary elements and $k=24$ on a single GPU. In this case, 
the setup time of the hierarchical matrix consumes about 40 seconds. A single
iteration of the CG solver requires in average 0.035 seconds with a total
of about 100 iterations.

To further increase the problem size, we increased the number of GPUs
by a distributed-memory parallelization, cf.~Table~\ref{tab:runtimes}. Using 
our task-based parallelization, cf.~Section~\ref{sec:parallelization}, we are 
able to treat the cube geometry test case with about 1.5 million boundary 
elements on 128 GPUs for $k=48$. The $\Hm$-matrix setup phase consumes
about 31 minutes. Once the setup has been computed, the system of linear
equations can be solved with only 0.19 seconds per iteration and a
total runtime of about 75 seconds. That is, we do the linear solve with 
1.5 million boundary elements in way less than one and a half minutes.

We also tried to further increase the problem size. However, we are
not able to accomplish this since the amount of memory consumed
on a single GPU (due to our strategy of independently carrying out
the data structure setup on all GPUs) becomes too high for the $6$ GB
of GPU RAM of the Tesla K20X GPUs. The use of a more recent GPU
like the Tesla P100 with $16$ GB of GPU RAM would of course improve the
situation.
This GPU would also achieve a much higher
performance in the $\Hm$-matrix setup, since our model BEM code would
run much faster than on the rather old Tesla K20X cards. In the future,
we also aim at combining a domain-decomposition parallelization with
the task-based parallelization as for example in \cite{Bebendorf2008,
Kravcenko2018} to solve even much larger problem sizes.

\paragraph{Performance vs.~accuracy.}
As discussed in Section~\ref{sec:convergence}, the increase in the
number of terms utilized in the adaptive cross approximation has an
important impact on the accuracy of the approximate solution. Therefore,
we analyzed its impact on the performance for the cube geometry and the
gearwheel geometry test case. In the first test case, we fix the problem
size to about $400000$ boundary elements and increase the number of 
terms $k$ from $24$ to up to $160$. Note that the timings used in this 
paragraph correspond to a single test run instead of five test runs.
Table~\ref{tab:runtimes} collects the runtime results of this test case
in the second row block. From a theoretical point of view, we should
expect a quadratic increase in the $\Hm$-matrix setup runtime with respect to $k$. In practice,
this increase is visible in the $\Hm$-matrix setup (including the ACA
approximation of the admissible blocks) for smaller $k$. However,
for larger $k$, this increase becomes smaller. We assume that this
behavior is due to the batching of the linear algebra operations
involved in ACA. In fact, the larger the problem, the better it is
possible to hide GPU latencies. We observe a
much smaller runtime increase in the CG solver.

We tried the same experiment for the real-world gearwheel geometry
test case with a problem size of about 1.2 million boundary elements  
with results given in the fourth row block of Table~\ref{tab:runtimes}.
Here, no quadratic runtime increase is visible. Instead, runtime is
increased sub-linearly. In this test case, the runtime for the dense
treatment of non-admissible matrix blocks seems to dominate the runtime.
To summarize, an increase in the number of ACA terms has only a
rather small impact on the overall runtime of our implementation.

\paragraph{Speed-up efficiency}
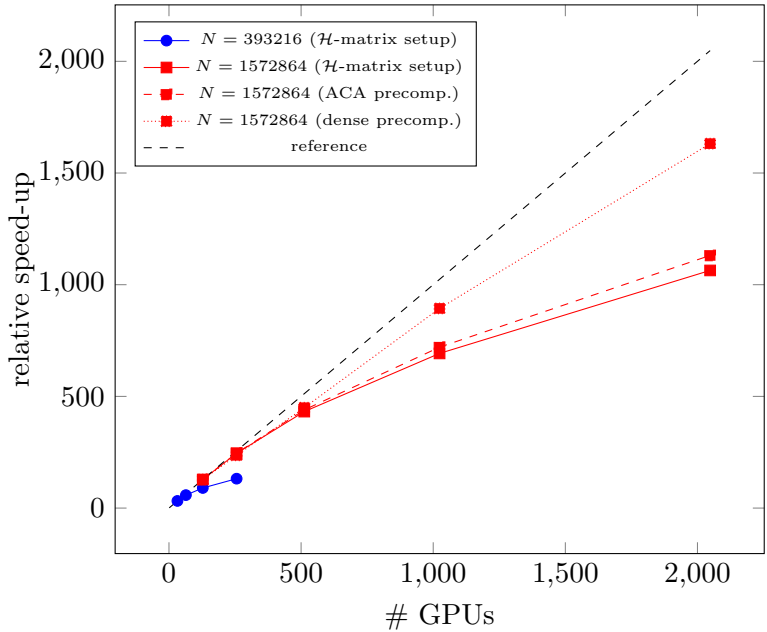
\begin{figure}[t]
\begin{tikzpicture}
\begin{axis}
[width=0.8\textwidth,
height=0.7\textwidth,
xlabel=\# GPUs,
ylabel=relative speed-up,
legend style={legend pos = north west, font=\tiny}]
\addplot[blue,mark=*] table{strong_scaling_cube_level_8.csv};
\addplot[red,mark=square*] table[x index=0,y index=1]{strong_scaling_cube_level_9.csv};
\addplot[red,mark=square*, dashed] table[x index=0,y index=2]{strong_scaling_cube_level_9.csv};
\addplot[red,mark=square*, densely dotted] table[x index=0,y index=3]{strong_scaling_cube_level_9.csv};
\addplot[black,dashed,domain=1:2048] {x};
\legend{{$N=393216$ ($\Hm$-matrix setup)}, {$N=1572864$ ($\Hm$-matrix setup)}, {$N=1572864$ (ACA precomp.)}, {$N=1572864$ (dense precomp.)}, {reference}}
\end{axis}
\end{tikzpicture}
\caption{\label{fig:strongScalingCube}The parallel speed-up efficiency
for $\Hm$-matrix setup of the cube geometry test case
\textit{(solid red line)} is above 67 percent on up to 1024 GPU
(starting form 128 GPUs) and still reaches above 50 percent on 2048
GPUs. Focussing on the pre-computation of (dense) non-admissible blocks,
we even achieve almost 80 percent parallel speed-up efficiency.}
\end{figure}
One of the main goals of this work is the increase in performance
for the \textit{$\Hm$-matrix setup} by the use of more GPUs. To
showcase the achieved efficiency, we perform a parallel speed-up
/ strong scaling study for the cube and the gearwheel geometry.
The results for the cube geometry are given in
Figure~\ref{fig:strongScalingCube} and in the third row block of
Table~\ref{tab:runtimes}.

In Figure~\ref{fig:strongScalingCube}, we show the results of a
parallel speed-up analysis for problem sizes $N=393216$ and
$N=1572864$ with $k=48$. While the smaller problem size does
not scale on large GPU counts, we observe a decent result for
the larger test case. Here, a relative parallel speed-up efficiency
of more than $67$ percent is achieved starting from 128 GPUs 
and going to up to 1024 GPUs. The effective runtime of the 
$\Hm$-matrix setup on 1024 GPUs is less than $5.7$ minutes, while 
the solution time by the CG solver is less than 20 seconds. In total,
we therefore need on 1024 GPUs less than 6 minutes for the
$\Hm$-matrix setup and solve for a problem size of more than
1.5 million boundary elements.

When trying to further speed-up the problem, we still achieve
an acceptable speed-up efficiency of above 50 percent on 2048
GPUs. In addition, we looked more closely in the contribution
to the scalability of the different work loads in the $\Hm$-matrix 
setup. As shown in Figure~\ref{fig:strongScalingCube}, the 
pre-computation of the (dense) non-admissible blocks scales
much better than the approximation of the admissible blocks by
ACA. From Table~\ref{tab:runtimes}, we can depict that the
strong scaling efficiency of the dense computations is in the
range of 80 percent, even on 2048 GPUs. The scaling of the ACA
work load is only a little bit more efficient than the overall
$\Hm$-matrix setup phase. The overall setup phase is always less
efficiencient than the dense and ACA work loads, since the
GPU-parallel data structure setup, i.e.~the block cluster
tree traversal, is not parallelized in a distributed-memory
manner, compare Section~\ref{sec:parallelization}.

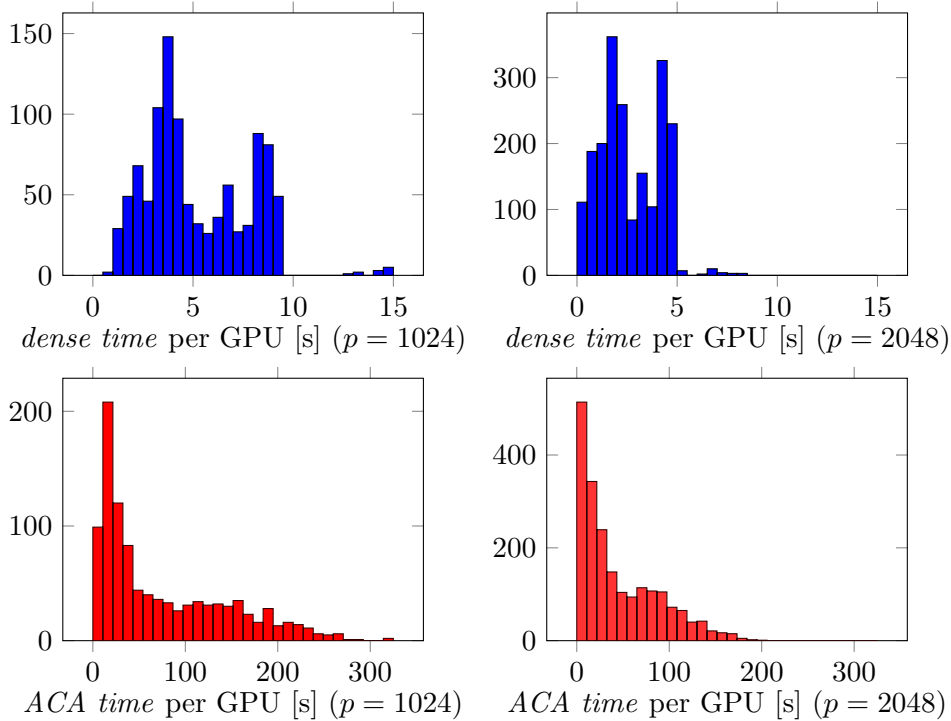
\begin{figure}[t]
\begin{tikzpicture}
\begin{axis}[
    width=0.5\textwidth,
    height=0.4\textwidth,
    ybar,
    ymin=0,
    xlabel={\textit{dense time} per GPU [s] ($p=1024$)}
]
\addplot +[
    black,
    fill=blue,
    hist={
        bins=30,
        data min=0,
        data max=15,
    }   
] table [y expr=\thisrowno{0}/1000] {precompute_dense_timings_level_9_1024_gpus.csv};
\end{axis}
\end{tikzpicture}
\begin{tikzpicture}
\begin{axis}[
    width=0.5\textwidth,
    height=0.4\textwidth,
    ybar,
    ymin=0,
    xlabel={\textit{dense time} per GPU [s] ($p=2048$)}
]
\addplot +[
    black,
    fill=blue,
    hist={
        bins=30,
        data min=0,
        data max=15,
    }   
] table [y expr=\thisrowno{0}/1000] {precompute_dense_timings_level_9_2048_gpus.csv};
\end{axis}
\end{tikzpicture}\\
\begin{tikzpicture}
\begin{axis}[
    width=0.5\textwidth,
    height=0.4\textwidth,
    ybar,
    ymin=0,
    xlabel={\textit{ACA time} per GPU [s] ($p=1024$)}
]
\addplot +[
    red!20!black,
    fill=red,
    hist={
        bins=30,
        data min=0,
        data max=325,
    }   
] table [y expr=\thisrowno{0}/1000] {precompute_aca_timings_level_9_1024_gpus.csv};
\end{axis}
\end{tikzpicture}
\begin{tikzpicture}
\begin{axis}[
    width=0.5\textwidth,
    height=0.4\textwidth,
    ybar,
    ymin=0,
    xlabel={\textit{ACA time} per GPU [s] ($p=2048$)}
]
\addplot +[
    red!20!black,
    fill=red!80!white,
    hist={
        bins=30,
        data min=0,
        data max=325,
    }   
] table [y expr=\thisrowno{0}/1000] {precompute_aca_timings_level_9_2048_gpus.csv};
\end{axis}
\end{tikzpicture}
\caption{\label{fig:loadBalancing}We analyze the load
balancing in the speed-up study of the $\Hm$-matrix setup
(cube geometry test case, $N=1572864$, $k=48$) with a
special focus on the per-GPU runtimes of the pre-computation
of the non-admissible blocks \textit{(blue)} and the per-GPU
runtimes of the approximation of the admissible blocks
\textit{(red)}.}
\end{figure}
In Section~\ref{sec:parallelization}, we also discussed the
issue of load balancing. Effectively, we chose the model
assumption that a roughly equal amount of batched matrix
entries that are applied in a batched matrix-vector product
will result in similar runtime performance. We check the quality
of this assumption by examining the distribution of computational
runtime over all GPUs for the dense matrix blocks and the ACA
matrix approximations. The results of this study for
$N=1572864$, $k=48$ and $p=1024, 2048$ are shown in
Figure~\ref{fig:loadBalancing} as histogram plots. Qualitatively,
the changes between 1024 and 2048 GPUs are only small, that is we
only consider $p=1024$. In practice, our model assumption does not
yet lead to an optimal load balancing. While a major part of the
timings for the dense matrix operations (see the blue histogram 
on the left-hand side in Figure~\ref{fig:loadBalancing}) is nicely
scattered around $5$ seconds, we have some non-optimal outliers
of up to 15 seconds. Nevertheless, the dense operations have only
a moderate influence on the overall scalability due to their small
maximum time. In contrast, the matrix block approximations 
by ACA (see the red histogram on the left-hand side in
Figure~\ref{fig:loadBalancing}) last up to more than 300 seconds.
We especially observe the rather large portion of GPUs that only need
a very small amount of time. In the future, we aim at improving
the multi-GPU load balancing by techniques proposed
e.g.~in~\cite{Bebendorf2008,Kravcenko2018}. However, while these
techniques work well in the context of non-batched operations, we
assume that their combination with batching will still be
sub-optimal on GPUs. Therefore, further research has to be
carried out in order to improve the load balancing.

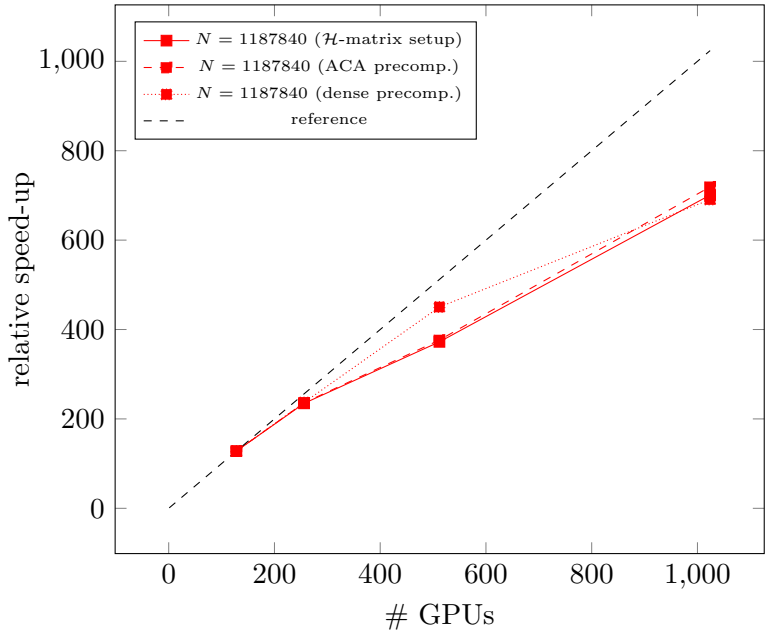
\begin{figure}[t]
\begin{tikzpicture}
\begin{axis}
[width=0.8\textwidth,
height=0.7\textwidth,
xlabel=\# GPUs,
ylabel=relative speed-up,
legend style={legend pos = north west, font=\tiny}]
\addplot[red,mark=square*] table[x index=0,y index=1]{strong_scaling_gearwheel3_level_6.csv};
\addplot[red,mark=square*, dashed] table[x index=0,y index=2]{strong_scaling_gearwheel3_level_6.csv};
\addplot[red,mark=square*, densely dotted] table[x index=0,y index=3]{strong_scaling_gearwheel3_level_6.csv};
\addplot[black,dashed,domain=1:1024] {x};
\legend{{$N=1187840$ ($\Hm$-matrix setup)}, {$N=1187840$ (ACA precomp.)}, {$N=1187840$ (dense precomp.)}, {reference}}
\end{axis}
\end{tikzpicture}
\caption{\label{fig:strongScalingGearwheel}Considering the real-world
test case of a gearwheel geometry with $N=1187804$ and $k=24$, we
observe a relative parallel speed-up efficiency of above 68 percent
going from 128 to 1024 GPUs. This result is almost identical to the
much simpler cube geometry test case.}
\end{figure}
We finally repeat the parallel speed-up efficiency study for the gearwheel
geometry test case with $N=1187840$ and $k=24$. The speed-up results
are graphically displayed in Figure~\ref{fig:strongScalingGearwheel}.
In this test case, the speed-up for the dense matrix work load and the
ACA approximation work load are well aligned to the overall $\Hm$-matrix
setup speed-ups. In total, we achieve above 68 percent of parallel speed-up
efficiency going from 128 GPUs to 1024 GPUs. This is a decent result.
Moreover, as shown by Table~\ref{tab:runtimes} in the last row block, the
runtime for the $\Hm$-matrix setup on 1024 is about $8.3$ minutes, while
the per-iteration runtime of the CG solver is about $0.06$ seconds with
a runtime of less than $29$ seconds for the full CG solve. That is, a
total of $8.8$ minutes is needed to solve a BEM problem with $1187840$
boundary elements on a real-world geometry on 1024 (rather old) GPUs, 
recalling that the implemented GPU BEM solver is only intended to be a 
model application for the underlying \texttt{hmglib} library.

\section{Conclusions}\label{sec:conclusions}
In this work, we considered the distributed-memory parallel multi-GPU
parallel solution of boundary integral equations by the hierarchical
matrix library \texttt{hmglib}. The main contribution of this work was
the extension and distributed-memory parallelization of \texttt{hmglib},
such that Galerkin matrices from boundary element method discretizations
given by arbitrary BEM codes can be approximated in a multi-GPU parallel
way. Our multi-GPU parallelization of the $\Hm$-matrix library uses a
task-based parallelization. Numerical studies and performance analysis
were carried out with a model GPU BEM solver for piecewise constant
boundary elements, which is based on an existing in-house CPU solver. 
This model GPU BEM solver was merely designed for a test of the \texttt{hmglib}
library, however not with the intention to compete with other BEM solvers
in the field. Our two numerical test cases showed, both, roughly 67 percent
relative parallel speed-up efficiency going from 128 to 1024 GPUs on
the GPU cluster \textit{Titan}. This is a decent speed-up result. A cube
geometry test case with about $1.5$ million boundary elements could be 
solved within less than 6 minutes (5.7 minutes for the setup, 20 seconds for
the CG solver) on 1024 GPUs. The real-world gearwheel geometry test case
with about $1.2$ million unknowns could be solved within $8.8$ minutes
on 1024 GPUs.

As future work, we consider the combination of a domain decomposition
parallelization with our task-based parallelization in order to scale to much
larger problem sizes. Such an approach should also allow to get a more
pronounced speed-up in the CG solver, for which we currently did not
aim for. Moreover, we plan to further investigate load balancing
techniques, which are assumed to have a strong impact on
pre-asymptotic runtimes.

\section*{Acknowledgements}
This work is funded by the Swiss National Science Foundation (SNF)
under project number $407540\_167186$. 
Furthermore, code development and benchmarking tasks in this
research were done on resources of the Oak Ridge Leadership
Computing Facility
at the Oak Ridge National Laboratory, which is supported by
the Office of Science of the U.S.~Department of Energy under
Contract No.~DE-AC05-00OR22725.
All funding and support is gratefully acknowledged.

\bibliographystyle{plain}
\bibliography{bibliography}

\end{document}